\documentclass[preprintnumbers,amsmath,amssymb,nofootinbib]{revtex4}

\usepackage{graphicx}
\usepackage{epsfig}
\usepackage{color}

\newcommand{\beq}{\begin{equation}}
\newcommand{\eeq}{\end{equation}}
\newcommand{\bea}{\begin{eqnarray}}
\newcommand{\eea}{\end{eqnarray}}
\newcommand{\bwd}{\begin{widetext}}
\newcommand{\ewd}{\end{widetext}}

\begin{document}

\title{A Symplectic Multi-Particle Tracking Model for Self-Consistent Space-Charge Simulation
}


\author{Ji Qiang}
\email{jqiang@lbl.gov}
\affiliation{Lawrence Berkeley National Laboratory, Berkeley, CA 94720, USA}

\begin{abstract}
Symplectic tracking is important in accelerator beam dynamics simulation. 
So far, to the best of our knowledge, there is no self-consistent symplectic space-charge tracking
model available in the accelerator community.
In this paper, we present a two-dimensional and a three-dimensional symplectic multi-particle spectral model for space-charge tracking simulation. This model includes both the effect from external fields and the effect of self-consistent space-charge fields using a split-operator method. Such a model 
preserves the phase space structure and shows much less 
numerical emittance growth than the particle-in-cell model in
the illustrative examples.
\end{abstract}

\maketitle

\section{Introduction}

In high intensity accelerators, the nonlinear space-charge effect from charged
particle interactions inside the beam has significant impact on beam dynamics 
through the accelerator. It causes beam emittance growth, halo formation, and even 
particle losses along the accelerator.
To study the space-charge effect, multi-particle tracking has been employed to dynamically follow
those charged particles through the accelerator.
In the accelerator community, most of those multi-particle tracking codes
use particle-in-cell (PIC) method to include the space-charge effect self-consistently
in the simulation~\cite{friedman,parmteq,takeda,track,impact,toutatis,qin0,impact-t,amundson,GPT,opal}.

The particle-in-cell method is an efficient method in handling the space-charge effect self-consistently.
It uses a computational grid to obtain the charge density distribution from
a finite number of macroparticles and solves the Poisson equation on the grid at each 
time step. The computational cost is linearly 
proportional to the number of macroparticles, which makes the simulation fast for
many applications. However, those grid based, momentum conserved, PIC codes do not 
satisfy the symplectic condition of classic multi-particle dynamics. 
Violating the symplectic condition
in multi-particle tracking might not be an issue in a single pass system such as
a linear accelerator. In a circular accelerator, 
violating the symplectic condition
may result in undesired numerical errors in the long-term tracking simulation. 
This issue together with the numerical grid heating was brought up during
the 2015 space-charge workshop at Oxford~\cite{sc2015}. 
A gridless spectral based macroparticle model 
was suggested by the author at the workshop to mitigate the numerical grid heating
and to satisfy the symplectic condition of particle tracking.

Multi-symplectic particle-in-cell model was
proposed to study Vlasov-Maxwell system and electrostatic system in plasmas using
a variational method~\cite{xiao,evstatiev,shadwick,qin,webb}.
To study the space-charge effect in high
intensity beams, a quasi-static model is normally employed. In the quasi-static
model, a moving beam frame is used to contain all charged particles through
the accelerator. 
The Poisson equation is solved in the beam frame to obtain electric Coulomb fields from
the charged particles.
These electric fields are transformed to the laboratory frame through the Lorentz transformation.
The space-charge forces acting on each individual particle include both the electric fields
and the magnetic fields, which is different from the electrostatic model that includes only
electric fields.
To the best of our knowledge, at present,
there is no symplectic self-consistent space-charge model available in the accelerator community. 
In this paper, following the idea suggested at the Oxford workshop,
we present a two-dimensional and a three-dimensional symplectic quasi-static multi-particle
tracking model for space-charge simulations. The model presented here starts from the multi-particle
Hamiltonian directly and uses a gridless spectral method to calculate the space-charge forces. 

The organization of this paper is as follows: after the introduction, we present the
symplectic multi-particle tracking model including the space-charge effect in Section II; 
We present
a symplectic space-charge transfer map for a 2D coasting beam in Section III and a symplectic space-charge
map for a 3D bunched beam in Section IV; We discuss computational complexity of the proposed model in
Section V and draw conclusions in Section VI.

\section{Symplectic Multi-Particle Tracking With Space-Charge Effects}

In the accelerator beam dynamics simulation, for a multi-particle system with $N_p$ charged particles subject to both a space-charge self field
and an external field, an approximate Hamiltonian of the system can 
be written as~\cite{rob1,forest,alex}:
\begin{eqnarray}
	H & = & \sum_{i=1}^{N_p} {\bf p}_i^2/2 + \frac{1}{2} \sum_{i=1}^{N_p} 
	\sum_{j=1}^{N_p} q \varphi({\bf r}_i,{\bf r}_j)
	+ \sum_{i=1}^{N_p} q \psi({\bf r}_i)
	\label{htot}
\end{eqnarray}
where $H({\bf r}_1,{\bf r}_2,\cdots,{\bf r}_{N_p},{\bf p}_1,{\bf p}_2,\cdots,{\bf p}_{N_p};s)$ denotes the Hamiltonian of the system using distance $s$ as an independent variable, 
$\varphi$ is related to the space-charge interaction potential 
between the charged
particles $i$ and $j$ (subject to appropriate boundary conditions), $\psi$ denotes the
potential associated with the external field,
${\bf r}_i=(x_i,y_i,\theta_i=\omega \Delta t)$ denotes the normalized canonical
spatial coordinates of particle $i$, ${\bf p}_i=(p_{xi},p_{yi},p_{ti}=-\Delta E/mC^2)$ the normalized canonical momentum
coordinates of particle $i$, and $\omega$ the reference angular frequency,
$\Delta t$ the time of flight to location $s$, $\Delta E$ the energy
deviation with respect to the reference particle, $m$ the rest mass of the
particle, and $C$ the speed of light in vacuum.
The equations governing the motion of individual particle
$i$ follows the Hamilton's equations as:
\begin{eqnarray}
	\frac{d {\bf r}_i}{d s} & = & \frac{\partial H}{\partial {\bf p}_i} \\
	\frac{d {\bf p}_i}{d s} & = & -\frac{\partial H}{\partial {\bf r}_i} 
\end{eqnarray}
Let $\zeta$ denote a 6N-vector of coordinates,
the above Hamilton's equation can be rewritten as:
\begin{eqnarray}
	\frac{d \zeta}{d s} & = & -[H, \zeta] 
\end{eqnarray}
where [\ ,\ ] is the Poisson bracket. A formal solution for above equation
after a single step $\tau$ can be written as:
\begin{eqnarray}
	\zeta (\tau) & = & \exp(-\tau(:H:)) \zeta(0)
\end{eqnarray}
Here, we have defined a differential operator $:H:$ as $: H : g = [H, \ g]$, 
for arbitrary function $g$. 
For a Hamiltonian that can be written as a sum of two terms $H =  H_1 + H_2$, an approximate
solution to above formal solution can be written as~\cite{forest1}
\begin{eqnarray}
	\zeta (\tau) & = & \exp(-\tau(:H_1:+:H_2:)) \zeta(0) \nonumber \\
  & = & \exp(-\frac{1}{2}\tau :H_1:)\exp(-\tau:H_2:) \exp(-\frac{1}{2}\tau:H_1:) \zeta(0) + O(\tau^3)
\end{eqnarray}
Let $\exp(-\frac{1}{2}\tau :H_1:)$ define a transfer map ${\mathcal M}_1$ and
$\exp(-\tau:H_2:)$ a transfer map ${\mathcal M}_2$, 
for a single step, the above splitting results in a second order numerical integrator
for the original Hamilton's equation as:
\begin{eqnarray}
	\zeta (\tau) & = & {\mathcal M}(\tau) \zeta(0) \nonumber \\
    & = & {\mathcal M}_1(\tau/2) {\mathcal M}_2(\tau) {\mathcal M}_1(\tau/2) \zeta(0)
	+ O(\tau^3)
	\label{map}
\end{eqnarray}
Using the above transfer maps ${\mathcal M}_1$ and  ${\mathcal M}_2$, a fourth order numerical integrator
can also be constructed as~\cite{forest1}:
\begin{eqnarray}
{\mathcal M}(\tau) & = & {\mathcal M}_1(\frac{s}{2}) {\mathcal M}_2(s) {\mathcal M}_1(\frac{\alpha s}{2}) {\mathcal M}_2((\alpha-1)s) {\mathcal M}_1(\frac{\alpha s}{2}) {\mathcal M}_2(s) {\mathcal M}_1(\frac{s}{2}) + O(\tau^5)
	\label{map4}
\end{eqnarray}
where $\alpha = 1-2^{1/3}$, and $s=\tau/(1+\alpha)$.
An even higher order accuracy integrator can be obtained following Yoshida's approach~\cite{yoshida}.
Assume that ${\mathcal M}_{2n}$ denotes a transfer map with an accuracy of order $2n$, the tranfer map
${\mathcal M}_{2n+2}$ with $(2n+2)$th order of accuracy can be obtained from the recursion equation:
\begin{eqnarray}
	{\mathcal M}_{2n+2}(\tau) & = & {\mathcal M}_{2n}(z_0\tau) {\mathcal M}_{2n}(z_1\tau) {\mathcal M}_{2n}(z_0 \tau) + O(\tau^{2n+3})
	\label{map2n}
\end{eqnarray}
where $z_0 = 1/(2-2^{1/(2n+1)})$ and $z_1 = -2^{1/(2n+1)}/(2-2^{1/(2n+1)})$.

The above numerical integrator Eqs.~\ref{map}-\ref{map2n} will be symplectic if both the transfer map
${\mathcal M}_1$ and the transfer map ${\mathcal M}_2$ are symplectic.
A transfer map ${\mathcal M}_i$ is symplectic if and only if
the Jacobian matrix $M_i$ of the transfer
map ${\mathcal M}_i$ satisfies the following condition:
\begin{eqnarray}
M_i^T J M_i = J 
\label{symp}
\end{eqnarray}
where $J$ denotes the $6N \times 6N$ matrix given by:
\begin{equation}
	J  =   \left( \begin{array}{cc}
			0 & I \\
			-I & 0
	\end{array} \right)
\end{equation}
and $I$ is the $3N\times 3N$ identity matrix.

For the given Hamiltonian in Eq.~\ref{htot}, we can choose $H_1$ as:
\begin{eqnarray}
	H_1 & = & \sum_{i=1}^{N_p} {\bf p}_i^2/2 + \sum_{i=1}^{N_p} q \psi({\bf r}_i)
\end{eqnarray}
A single charged particle magnetic optics method can be used to find a symplectic transfer map ${\mathcal M}_1$
for this Hamiltonian with the external fields from
most accelerator beam line elements~\cite{rob1,alex,mad}.


We can choose $H_2$ as:
\begin{eqnarray}
	H_2 & = & \frac{1}{2} \sum_{i=1}^{N_p} \sum_{j=1}^{N_p} q \varphi({\bf r}_i,{\bf r}_j)
\end{eqnarray}
which includes the space-charge effect and is only a function of positions. 
For the space-charge Hamiltonian $H_2({\bf r})$, the single
step transfer map ${\mathcal M}_2$ can be written as:
\begin{eqnarray}
	{\bf r}_i(\tau) & = & {\bf r}_i(0) \\
	{\bf p}_i(\tau) & = & {\bf p}_i(0) - \frac{\partial H_2({\bf r})}{\partial {\bf r}_i} \tau
	\label{map2}
\end{eqnarray}
The Jacobi matrix of the above 
 transfer map
${\mathcal M}_2$ is 
\begin{equation}
	M_2  =   \left( \begin{array}{cc}
			I & 0 \\
			L & I
	\end{array} \right)
\end{equation}
where $L$ is a $3N \times 3N$ matrix.
For $M_2$ to satisfy the symplectic condition Eq.~\ref{symp}, 
the matrix $L$ needs
to be a symmetric matrix, i.e.
\begin{equation}
	L = L^T
\end{equation}
Given the fact that $L_{ij} = \partial {\bf p}_i(\tau)/\partial {\bf r}_j =  - \frac{\partial^2 H_2({\bf r})}{\partial {\bf r}_i \partial {\bf r}_j} \tau $, the matrix $L$ will be symmetric as long as it 
is {\bf \it analytically calculated}
from the function $H_2$. This is also called jolt-factorization in nonlinear
single particle beam dynamics study~\cite{forest3}.
If both the transfer map
${\mathcal M}_1$ and the transfer map ${\mathcal M}_2$ 
are symplectic, the numerical integrator Eqs.~\ref{map}-\ref{map2n} for multi-particle tracking will be symplectic. In the following sections, 
we will derive the symplectic space-charge transfer map of $H_2$ for 
a two-dimensional coasting beam and for a three-dimensional bunched beam.

\section{Symplectic space-charge map for a coasting beam}

In a coasting beam, the Hamiltonian $H_2$ can be written as~\cite{rob1}:
\begin{eqnarray}
	H_2 & = & \frac{K}{2} \sum_{i=1}^{N_p} \sum_{j=1}^{N_p} \varphi({\bf r}_i,{\bf r}_j)
	\label{htot2}
\end{eqnarray}
where $K = q I/(2\pi \epsilon_0 p_0 v_0^2 \gamma_0^2)$ is the generalized perveance,
$I$ is the beam current, $\epsilon_0$ is the dielectric 
constant in vacuum, $p_0$ is the momentum of the reference
particle, $v_0$ is the speed of the reference particle, $\gamma_0$ is
the relativistic factor of the reference particle, and $\varphi$ is the 
space charge Coulomb interaction potential.
In this Hamiltonian, the effects of the direct electric potential and the
longitudinal vector potential are combined together.
The electric Coulomb potential in the Hamiltonian $H_2$ can be obtained 
from the solution of the Poisson equation.
In the following, we assume that the coasting beam is inside a rectangular perfect conducting pipe.
In this case, the two-dimensional Poisson's equation can be written as:
\begin{equation}
\frac{\partial^2 \phi}{\partial x^2} +
\frac{\partial^2 \phi}{\partial y^2} = - 4 \pi \rho
\label{poi2d}
\end{equation}
where
$\phi$ is the electric potential, and $\rho$ is the particle
density distribution of the beam.

The boundary conditions for the electric potential inside the rectangular 
conducting pipe are:
\begin{eqnarray}
	\label{bc1}
\phi(x=0,y) & = & 0  \\
\phi(x=a,y) & = & 0  \\
\phi(x,y=0) & = & 0  \\
\phi(x,y=b) & = & 0  
	\label{boundary}
\end{eqnarray}
where $a$ is the horizontal width of the pipe and $b$ is the vertical width
of the pipe. 

Given the boundary conditions in Eqs.~\ref{bc1}-\ref{boundary}, the electric potential $\phi$ and the
source term $\rho$ can be approximated using two sine functions as~\cite{gottlieb,fornberg,boyd,qiang1,qiang2}:
\begin{eqnarray}
	\rho(x,y)  = \sum_{l=1}^{N_l}\sum_{m=1}^{N_m} \rho^{lm} \sin(\alpha_l x) \sin(\beta_m y) \\
	\phi(x,y)  =  \sum_{l=1}^{N_l}\sum_{m=1}^{N_m} \phi^{lm} \sin(\alpha_l x) \sin(\beta_m y) 
\end{eqnarray}
where
\begin{eqnarray}
\label{rholm}
\rho^{lm}  = \frac{4}{ab}\int_0^a\int_0^b \rho(x,y) \sin(\alpha_l x) \sin(\beta_m y) \ dx dy \\
\phi^{lm}  = \frac{4}{ab}\int_0^a\int_0^b \phi(x,y) \sin(\alpha_l x) \sin(\beta_m y) \ dx dy
\end{eqnarray}
where $\alpha_l=l\pi/a$ and $\beta_m = m \pi/b$.
The above approximation
follows the numerical spectral Galerkin method since each basis function
satisfies the boundary conditions on the wall~\cite{gottlieb,boyd,fornberg}. 
For a smooth function,
this spectral approximation has an accuracy whose numerical error
scales as $O(\exp(-cN))$ with 
$c>0$, where $N$ is the number of the basis function (i.e. mode number in each
dimension) used in the approximation.
By substituting above expansions into the
Poisson Eq.~\ref{poi2d} and making use of the orthonormal condition of the sine functions,
we obtain
\begin{eqnarray}
	\phi^{lm} & = & \frac{4 \pi \rho^{lm}}{\gamma_{lm}^2}
	\label{odelm}
\end{eqnarray}
where $\gamma_{lm}^2 = \alpha_l^2 + \beta_m^2$. 

In the multi-particle tracking, the particle distribution function $\rho(x,y)$ can be represented as:
\begin{eqnarray}
	\rho(x,y) = \frac{1}{N_p}\sum_{j=1}^{N_p} \delta(x-x_j)\delta(y-y_j)
\end{eqnarray}
where 
$\delta$ is the Dirac function.
Using the above equation and Eq.~\ref{rholm} and Eq.~\ref{odelm}, we obtain:
\begin{eqnarray}
	\phi^{lm}  = \frac{4 \pi} {\gamma_{lm}^2}\frac{4}{ab} \frac{1}{N_p} \sum_{j=1}^{N_p} \sin(\alpha_l x_j) \sin(\beta_m y_j)
\end{eqnarray}
and the electric potential as:
\begin{eqnarray}
	\phi(x,y)  = {4 \pi} \frac{4}{ab} \frac{1}{N_p} \sum_{j=1}^{N_p} 
	\sum_{l=1}^{N_l} \sum_{m=1}^{N_m} 
	\frac{1}{\gamma_{lm}^2} \sin(\alpha_l x_j) \nonumber \\
	\sin(\beta_m y_j) \sin(\alpha_l x) \sin(\beta_m y)
\end{eqnarray}
From the above electric potential, the interaction potential 
$\varphi$ between particles $i$ and $j$ can be
written as:
\begin{eqnarray}
	\varphi(x_i,y_i,x_j,y_j)  = {4 \pi} \frac{4}{ab} \frac{1}{N_p}  
	\sum_{l=1}^{N_l} \sum_{m=1}^{N_m} 
	\frac{1}{\gamma_{lm}^2} \sin(\alpha_l x_j) \nonumber \\
	\sin(\beta_m y_j) \sin(\alpha_l x_i) \sin(\beta_m y_i)
\end{eqnarray}
Now, the space-charge Hamiltonian $H_2$ can be written as:
\begin{eqnarray}
	H_2  = 4\pi \frac{K}{2}\frac{4}{ab} \frac{1}{N_p} \sum_{i=1}^{N_p} \sum_{j=1}^{N_p} \sum_{l=1}^{N_l} \sum_{m=1}^{N_m} 
	\frac{1}{\gamma_{lm}^2} \sin(\alpha_l x_j) \nonumber \\
	\sin(\beta_m y_j) \sin(\alpha_l x_i) \sin(\beta_m y_i)
\end{eqnarray}
The one-step symplectic transfer map ${\mathcal M}_2$ of the 
particle $i$ with this Hamiltonian is given as:
\begin{eqnarray}
	p_{xi}(\tau) & = & p_{xi}(0) -
	\tau 4\pi {K}\frac{4}{ab} \frac{1}{N_p} \sum_{j=1}^{N_p} 
	\sum_{l=1}^{N_l} \sum_{m=1}^{N_m} 
	\frac{\alpha_l}{\gamma_{lm}^2} \nonumber \\
& & \sin(\alpha_l x_j) \sin(\beta_m y_j) \cos(\alpha_l x_i) \sin(\beta_m y_i)  \nonumber \\
	p_{yi}(\tau) & = & p_{yi}(0) -
	\tau 4\pi {K}\frac{4}{ab} \frac{1}{N_p} \sum_{j=1}^{N_p} 
	\sum_{l=1}^{N_l} \sum_{m=1}^{N_m} 
	\frac{\beta_m}{\gamma_{lm}^2} \nonumber \\
& & \sin(\alpha_l x_j) \sin(\beta_m y_j) \sin(\alpha_l x_i) \cos(\beta_m y_i)
\end{eqnarray}
Here, both $p_{xi}$ and $p_{yi}$ are normalized by the reference particle momentum $p_0$.
Using the symplectic transfer map ${\mathcal M}_1$ for the external field Hamiltonian 
$H_1$ from an optics code and the transfer map ${\mathcal M}_2$,
one obtains a 
symplectic multi-particle
tracking model including the self-consistent space-charge effect
following Eqs.~\ref{map}-\ref{map2n}. 

\begin{figure}[!htb]
   \centering
   \includegraphics*[angle=270,width=174pt]{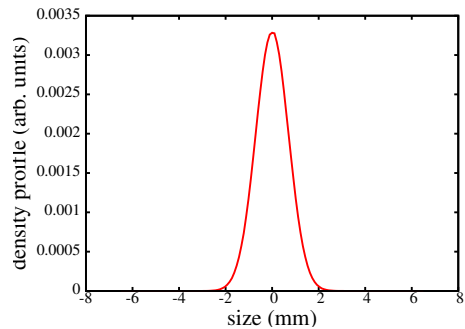}
   \caption{Charge density distribution along the $x$ axis.}
   \label{dist}
\end{figure}
\begin{figure}[!htb]
   \centering
   \includegraphics*[angle=270,width=174pt]{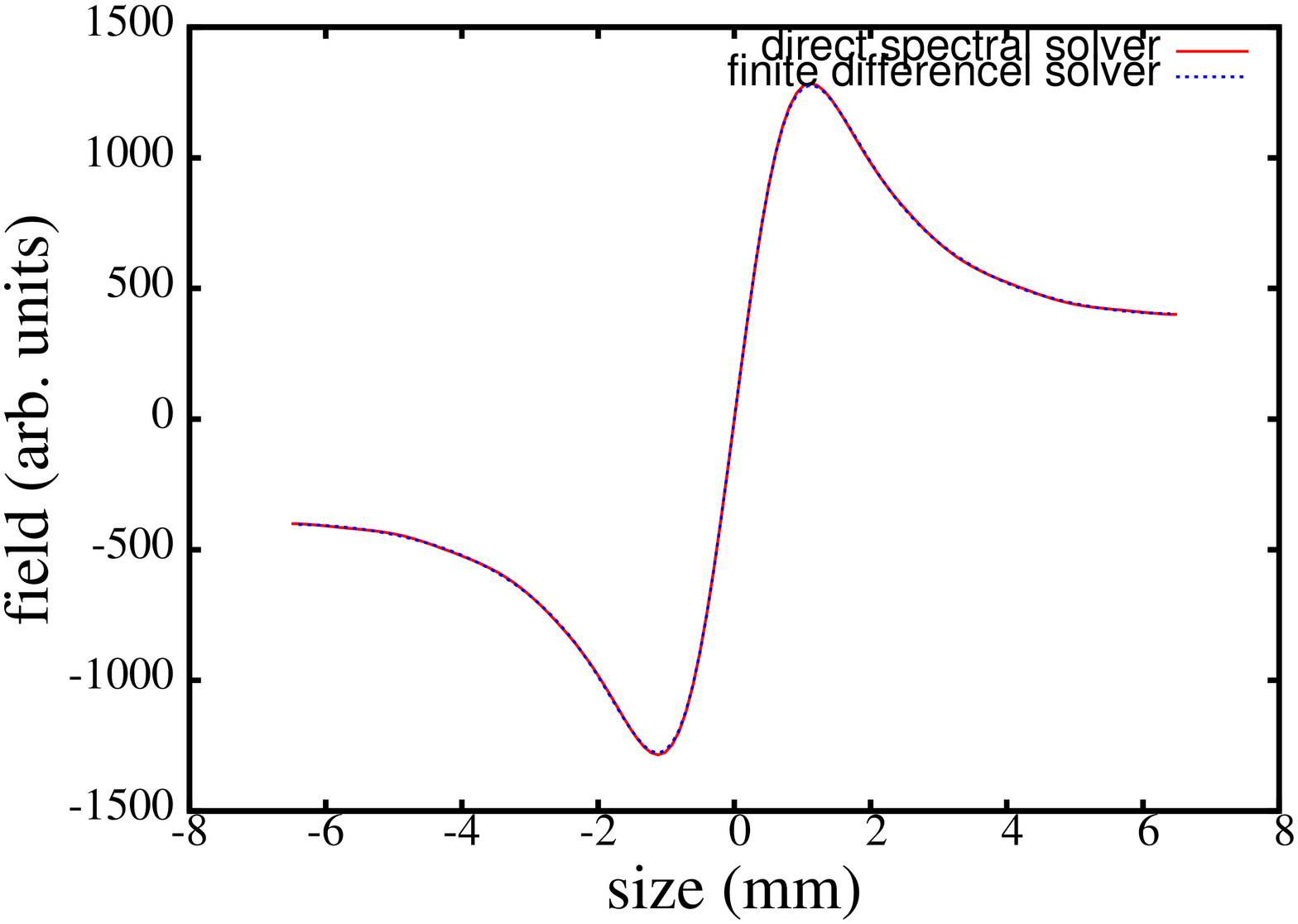}
   \includegraphics*[angle=270,width=174pt]{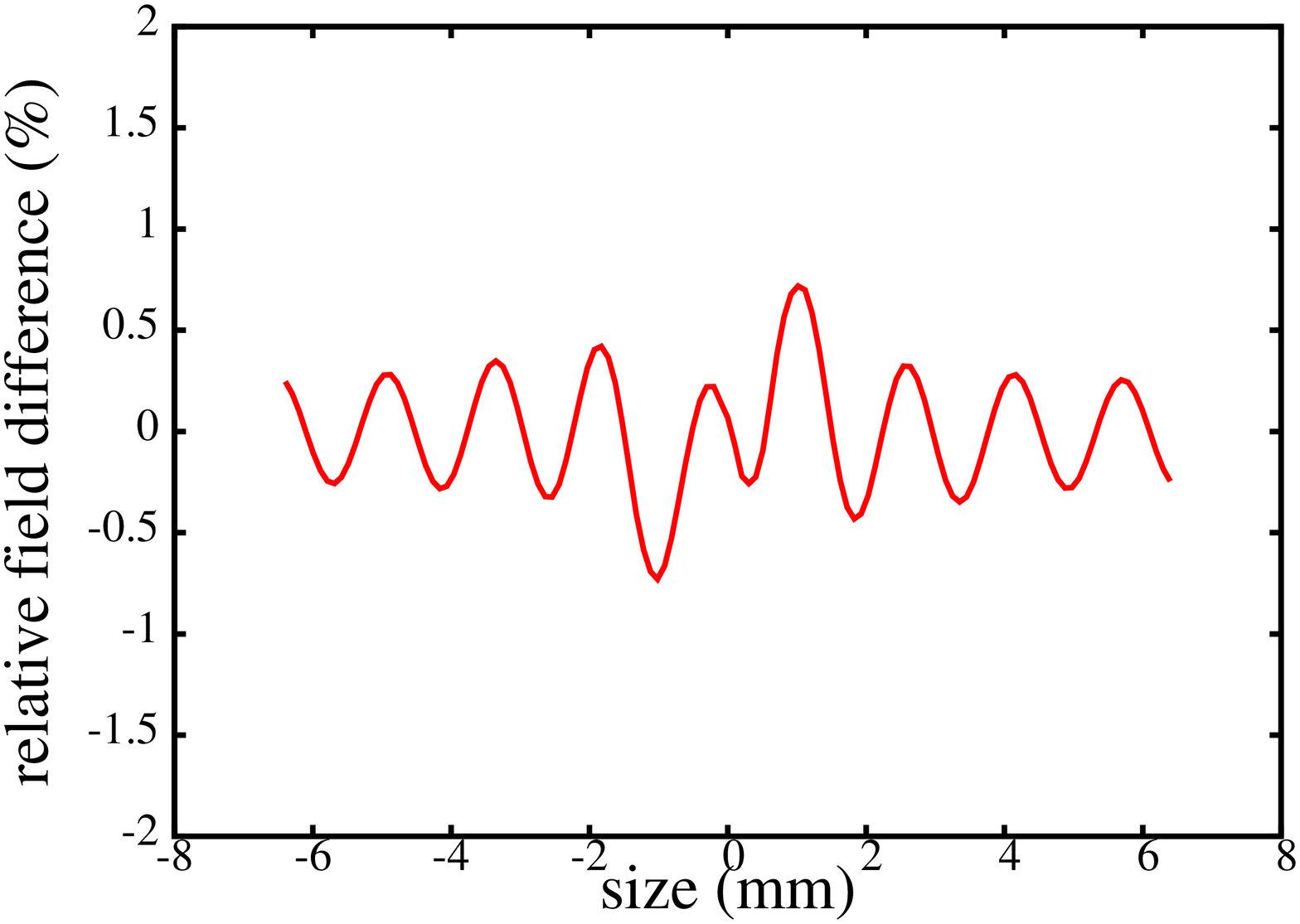}
   \caption{Electric field on $x$ axis from the above direct spectral solver
   (red) and from the 2nd order finite difference solver (green) (left), and the normalized relative field
   difference (right).}
   \label{field}
\end{figure}

As an illustration of above symplectic multi-particle tracking model,
we simulated a $1$ GeV coasting proton beam transporting through a rectangular
perfect conducting pipe with a FODO lattice for transverse focusing.
%
The initial transverse density distribution is assumed to be
a Gaussian function given in Fig.~\ref{dist}. 
We computed the electric field along the $x$ axis using the above direct gridless 
spectral solver with $15\times 15$ modes and the electric field from
a second order finite difference solver with $129\times 129$ grid points. 
The results are shown in Fig.~\ref{field}.
The solution of the spectral solver agrees with
that of the finite difference solver very well with even $15\times 15$ modes
due to the fast convergence property of the spectral method. 
The relative maximum field difference (normalized by the maximum field amplitude)
between two solutions is below $2\%$.
The use of the mode number in this example is somewhat empirical. 
It depends on the physical problem to be solved. 
If one knows about the smallest spatial structure of the problem, 
one can choose the mode number with the wavelength to resolve this spatial 
structure. Without knowing the detailed structure in the particle density 
distribution, one can use a trial-and-error method until the appropriate
solution is attained. 

\begin{figure}[!htb]
   \centering
   \includegraphics*[angle=270,width=174pt]{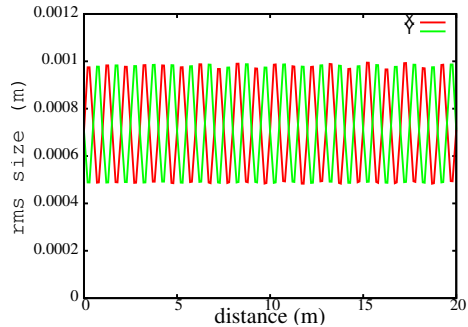}
   \caption{RMS envelope evolution of the beam.}
   \label{rms}
\end{figure}

Figure~\ref{rms} shows the proton beam root-mean-square (rms) envelope evolution through $20$ FODO lattice
periods. 
The FODO lattice used in this example consists of two quadrupoles and three
drifts in a single period. 
The total length of the period is $1$ meter. The zero current phase advance is about $87$ degrees and
the phase advance with $100$ A current is about $74$ degrees.
The relatively low intensity beam used in this example is to avoid space-charge driven resonance and to separate the numerical emittance growth from the physical emittance growth in the simulation.

\begin{figure}[!htb]
    \centering
    \includegraphics*[angle=270,width=70mm]{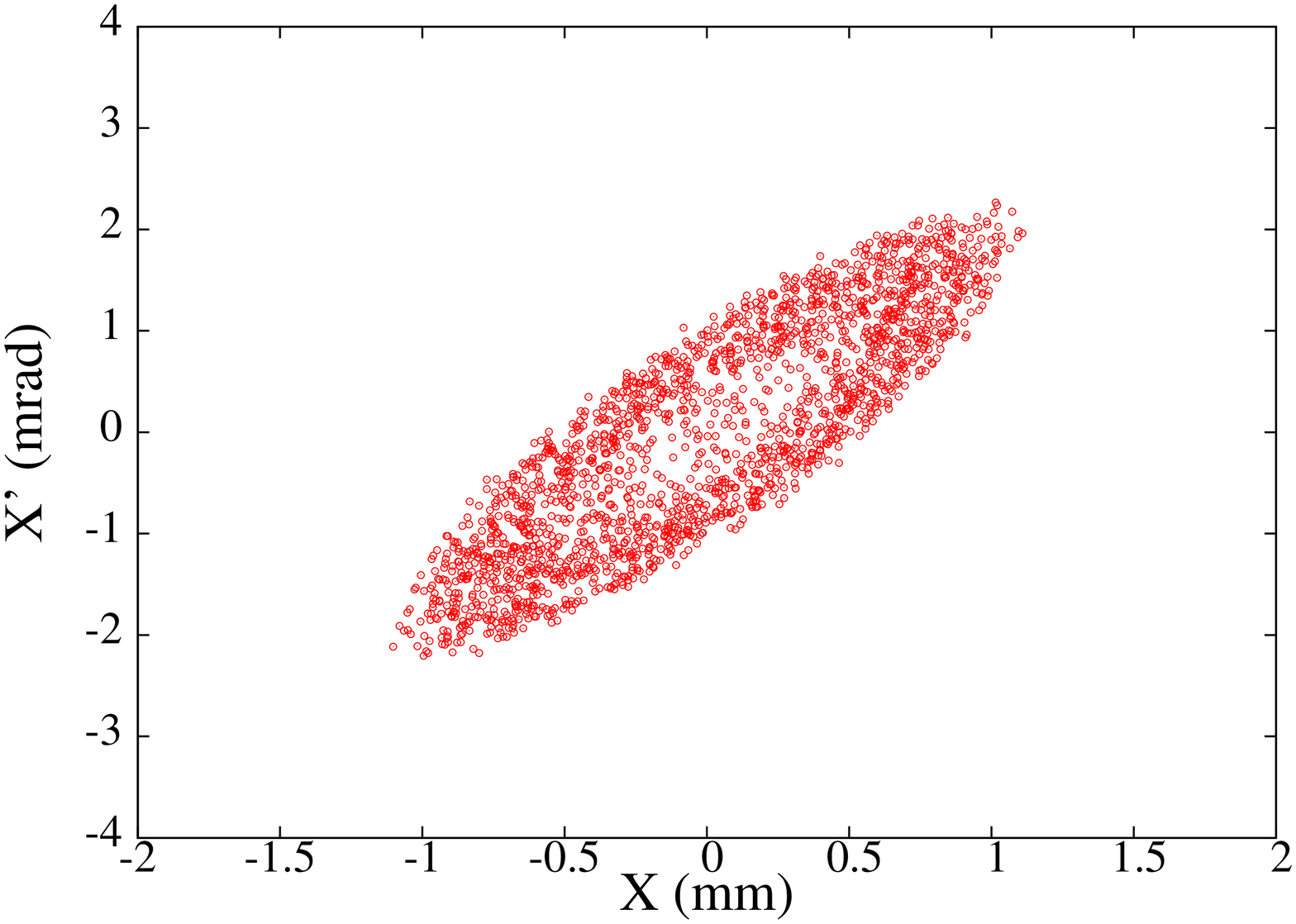}
    \includegraphics*[angle=270,width=70mm]{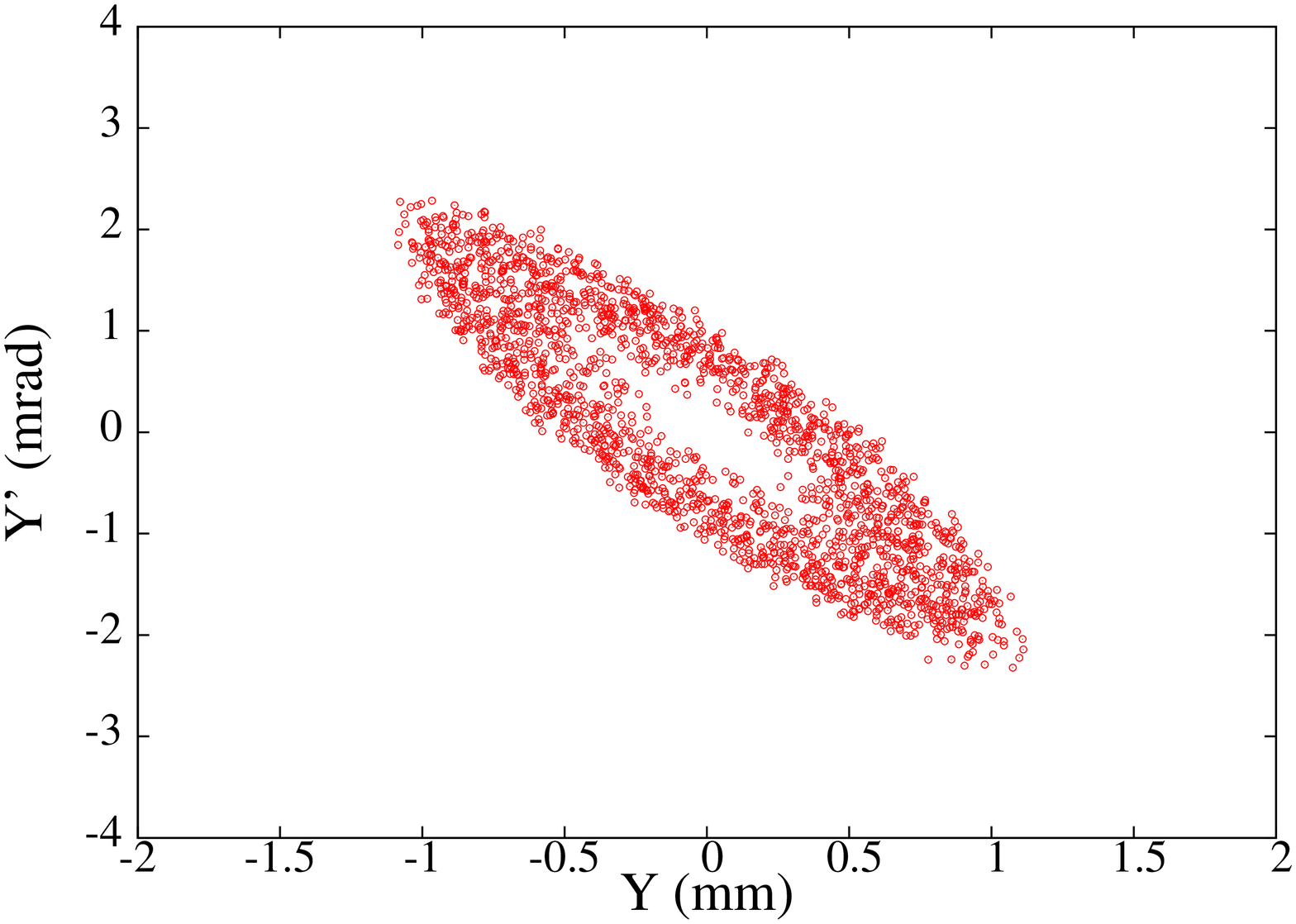}
    \includegraphics*[angle=270,width=70mm]{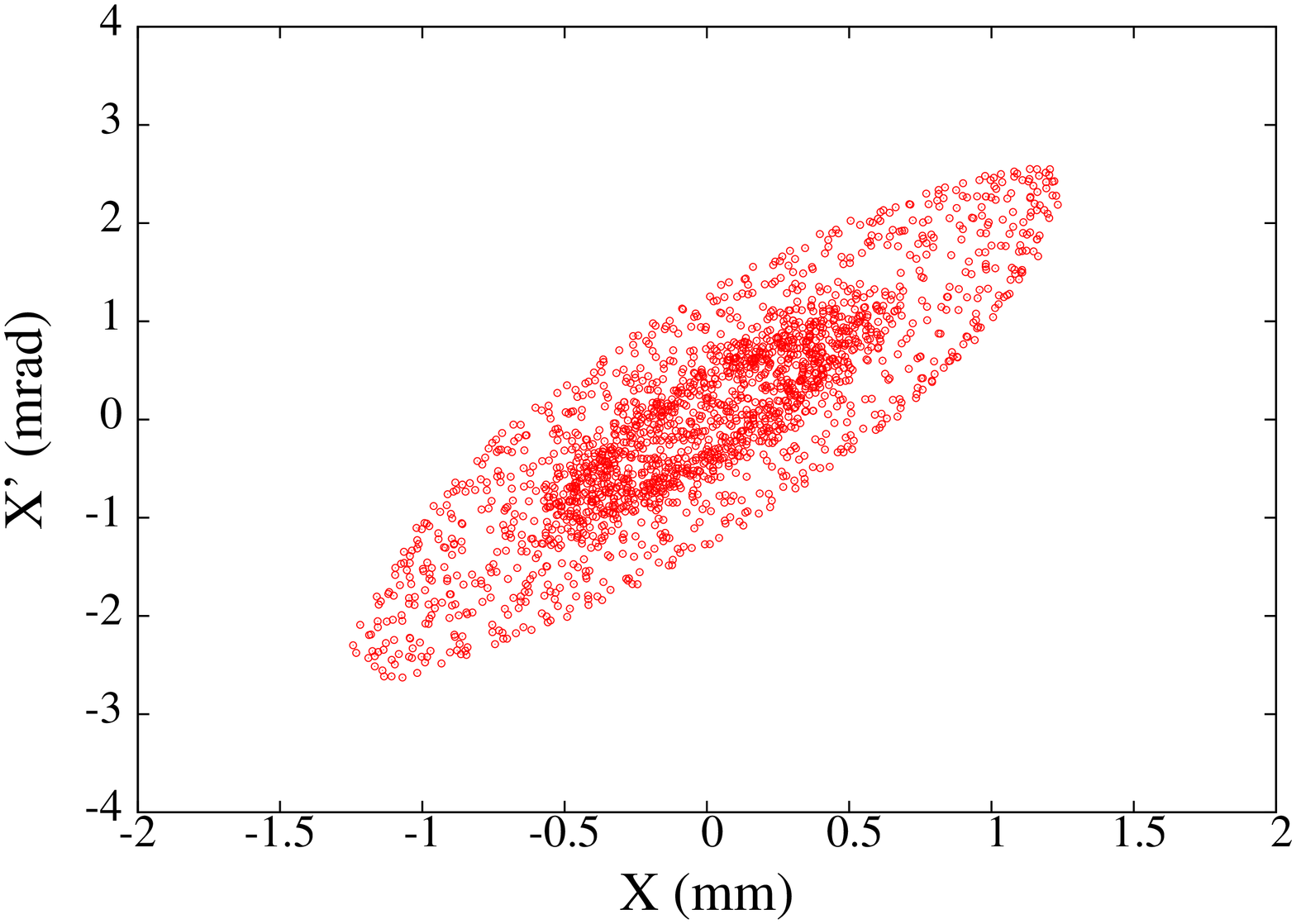}
    \includegraphics*[angle=270,width=70mm]{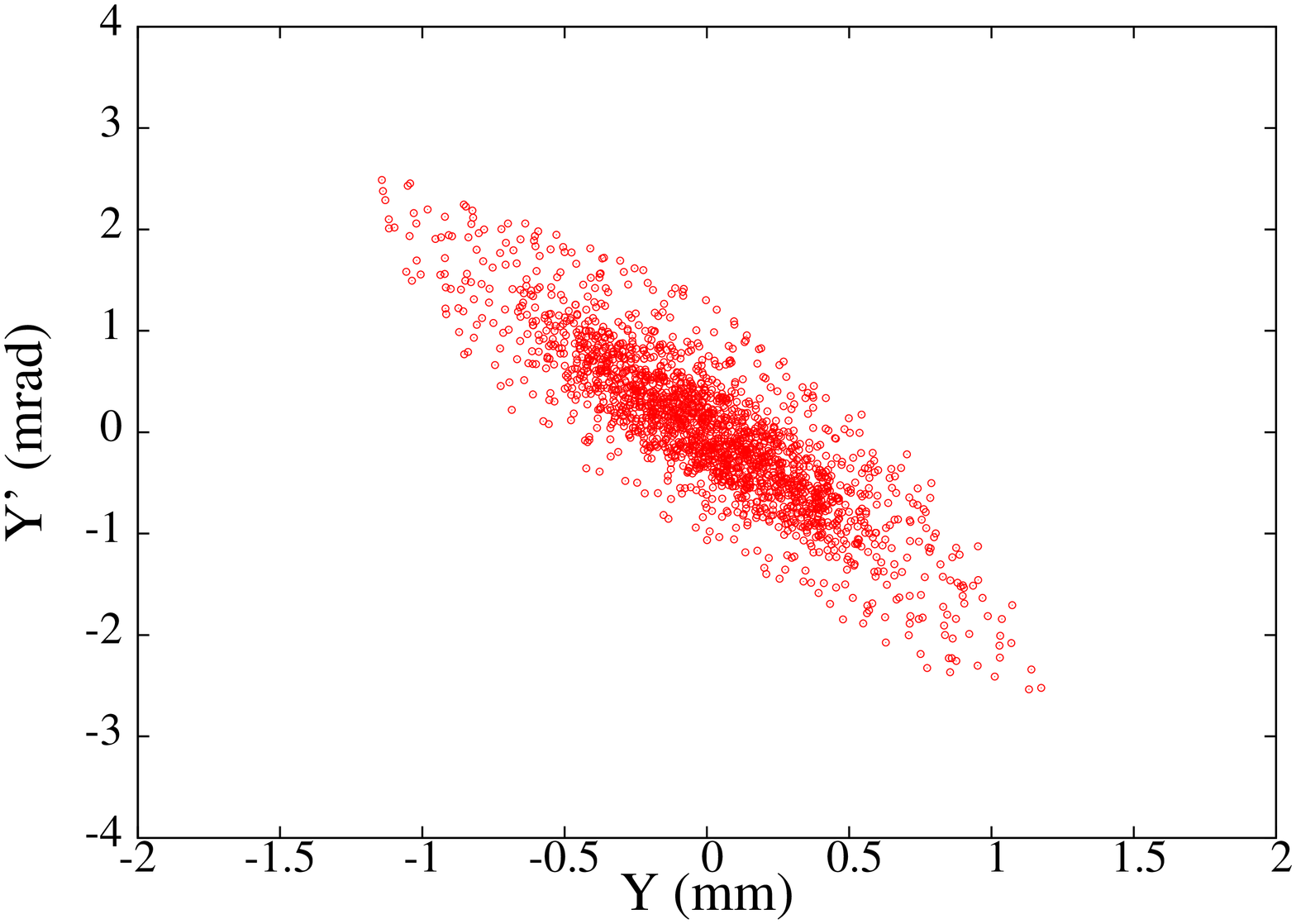}
    \caption{Stroboscopic plot (every 10 periods) of phase space evolution of 
	    a test particle from the symplectic-spectral model (top) and from the
	    PIC-finite difference model (bottom). }
    \label{phase2}
\end{figure}
\begin{figure}[!htb]
   \centering
   \includegraphics*[angle=270,width=174pt]{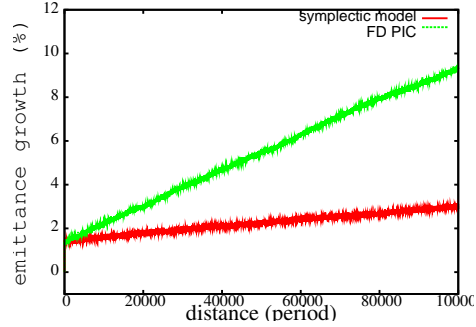}
   \caption{Four dimensional emittance growth evolution from the symplectic
   multi-particle spectral model (red) and from the PIC finite difference model(green).}
   \label{emtcmp1}
\end{figure}

The symplectic integrator is good for long term tracking since it
helps preserve phase space structure during the numerical integration.
Figure~\ref{phase2} shows the stroboscopic plots (every 10 periods) of 
$x-p_x$ and $y-p_y$ phase space evolution of a test particle through the
last $20,000$ periods of the total $100,000$
lattice periods including the self-consistent space-charge forces. 
As a comparison, we also show in this figure the phase
space evolution of the same initial test particle using the standard 
momentum conserved PIC method and the second order
finite difference solver for space-charge calculation~\cite{rob1,hockney}.
Qualitatively, these two models show similar shapes in phase space. 
However, looking into the details of the phase space, they
have quite different structures. The single particle phase space
from the PIC model shows a dense core while the phase space from the symplectic multi-particle
model shows a nearly hollow core.
Figure~\ref{emtcmp1} shows the 4-dimensional emittance growth 
($\frac{\epsilon_x}{\epsilon_{x0}}\frac{\epsilon_y}{\epsilon_{y0}}-1$)\% evolution from the
symplectic model and that from the PIC model.
It is seen that the symplectic model has a much smaller emittance growth
than the PIC model. This emittance growth
is a numerical artifact due to the small number of macroparticles
($50,000$) used in the simulation, which was studied in 
references~\cite{noise1,noise2,noise3}. The small number of macroparticles
introduces numerical errors in the computing of the electric potential
and results in the artificial emittance growth. 
A more detailed study of the numerical emittance growth associated with this new method is under way and will be reported in future publication. 
The apparent non-zero emittance growth at the beginning 
is due to the charge redistribution of the initial Gaussian distribution 
within a much shorter time scale (not visible in the plot) compared with the total plotting time scale of $100,000$ periods.

\section{Symplectic space-charge map for a 3D bunched beam}

In a 3D bunched beam, the Hamiltonian $H_2$ can be written as~\cite{rob2}:
\begin{eqnarray}
	H_2 & = &  \frac{\kappa \gamma_0}{2} \sum_{i=1}^{N_p} \sum_{j=1}^{N_p}
	\varphi({\bf r}_i,{\bf r}_j)
	\label{htot3}
\end{eqnarray}
where $\kappa = q/(lmC^2\gamma_0^2 \beta_0)$, $l=C/\omega$ is the scaling length, 
and $\beta_0 = v_0/C$. The
above Hamiltonian includes both the electric potential
and the longitudinal magnetic vector potential. The electric potential in 
the beam frame
can be obtained from the solution of a three-dimensional Poisson's
equation:
\begin{equation}
	\label{3dpoi}
\frac{\partial^2 \phi}{\partial x^2} +
\frac{\partial^2 \phi}{\partial y^2} +
\frac{\partial^2 \phi}{\partial z^2}
= - \frac{\rho}{\epsilon_0}
\end{equation}
where
$\rho$ is the charge
density distribution in the beam frame. 
The boundary conditions for the electric potential inside the rectangular 
perfect conducting pipe are:
\begin{eqnarray}
	\label{bc2}
\phi(x=0,y,z) & = & 0  \\
\phi(x=a,y,z) & = & 0  \\
\phi(x,y=0,z) & = & 0  \\
\phi(x,y=b,z) & = & 0  \\
\phi(x,y,z = -\infty) & = & 0  \\
\phi(x,y,z = \infty) & = & 0  
	\label{boundary2}
\end{eqnarray}
where $a$ is the horizontal width of the pipe,  $b$ is the vertical width
of the pipe. The solution of the
3D Poisson equation subject to the above boundary conditions
was studied 
with several numerical methods~\cite{qiang1,qiang2}. 
To obtain a fast analytical solution of the electric potential
we use an artificial boundary
condition in this study. 
Here, the longitudinal open boundary condition is approximated
by a finite domain Dirichlet boundary condition:
\begin{eqnarray}
\phi(x,y,z = 0) & = & 0  \\
\phi(x,y,z = c) & = & 0  
	\label{bc3}
\end{eqnarray}
where $c$ is the length of the domain that is
large enough so that the electric potential goes to zero at both ends of the
domain. The choice of the length of the domain depends on how
fast the electric potential vanishes outside the beam. 
From the reference~\cite{qiang2}, we know that the solution of the
electric potential for each transverse mode can be written as:
\begin{eqnarray}
	\phi^{lm}(z) & = & \frac{1}{2 \gamma_{lm} \epsilon_0} \int_{-\infty}^{\infty}
	\exp(-\gamma_{lm}|z-z'|) \rho^{lm}(z') \ dz'
\end{eqnarray}
where $\gamma^2_{lm} = (l \pi/a)^2+(m \pi/b)^2$. This solution
decreases exponentially as a function of $z$ outside the beam.
This suggests that a short distance (in the unit of aperture size)
might be sufficient to have the electric potential approach to
zero.

Given the boundary conditions in Eqs.~\ref{bc2}-\ref{bc3}, the electric potential $\phi$ and the
source term $\rho$ can be approximated using three sine functions as:
\begin{eqnarray}
	\rho(x,y,z)  = \sum_{l=1}^{N_l}\sum_{m=1}^{N_m} \sum_{n=1}^{N_n} \rho^{lmn} \sin(\alpha_l x) \sin(\beta_m y) \sin(\gamma_n z)\\
	\phi(x,y,z)  =  \sum_{l=1}^{N_l}\sum_{m=1}^{N_m} \sum_{n=1}^{N_n} \phi^{lmn} \sin(\alpha_l x) \sin(\beta_m y) \sin(\gamma_n z)
\end{eqnarray}
where
\begin{eqnarray}
\label{rholm2}
	\rho^{lmn}  = \frac{8}{abc}\int_0^a\int_0^b \int_0^c \rho(x,y,z) \sin(\alpha_l x) \sin(\beta_m y) \sin(\gamma_n z) \ dx dy dz \\
\phi^{lmn}  = \frac{8}{abc}\int_0^a\int_0^b \int_0^c \phi(x,y,z) \sin(\alpha_l x) \sin(\beta_m y) \sin(\gamma_n z) \ dx dy dz
\end{eqnarray}
where $\alpha_l=l\pi/a$, $\beta_m = m \pi/b$, $\gamma_n = n \pi/c$.
Substituting the above expansions into the
Poisson Eq.~\ref{3dpoi} and making use of the orthonormal condition of the sine functions,
we obtain
\begin{eqnarray}
	\phi^{lmn} & = & \frac{\rho^{lmn}}{\epsilon_0 \Gamma_{lmn}^2}
	\label{odelm2}
\end{eqnarray}
where $\Gamma_{lmn}^2 = \alpha_l^2 + \beta_m^2 + \gamma_n^2$. 

In the multi-particle tracking, the charge density $\rho(x,y,z)$ can be represented as:
\begin{eqnarray}
	\rho(x,y,z) = \sum_{j=1}^{N_p} w \delta(x-x_j)\delta(y-y_j)\delta(z-z_j)
\end{eqnarray}
where $w$ is the charge weight of each individual particle and $\delta$ is the Dirac function. 
Using the above equation and Eq.~\ref{rholm2} and Eq.~\ref{odelm2}, we obtain:
\begin{eqnarray}
	\phi^{lmn}  = \frac{1} {\epsilon_0 \Gamma_{lmn}^2}\frac{8}{abc} w \sum_{j=1}^{N_p} \sin(\alpha_l x_j) \sin(\beta_m y_j) \sin(\gamma_n z_j)
\end{eqnarray}
and the electric potential as:
\begin{eqnarray}
	\phi(x,y,z)  = \frac{1} {\epsilon_0}\frac{8}{abc} w \sum_{j=1}^{N_p} 
	\sum_{l=1}^{N_l} \sum_{m=1}^{N_m} 
	\sum_{n=1}^{N_n} \frac{1}{\Gamma_{lmn}^2} \sin(\alpha_l x_j) 
	\sin(\beta_m y_j) \sin(\gamma_n z_j)\sin(\alpha_l x) \sin(\beta_m y) \sin(\gamma_n z)
\end{eqnarray}
From the above electric potential, we obtain the interaction potential
between particles $i$ and $j$ as:
\begin{eqnarray}
	\varphi(x_i,y_i,z_i,x_j,y_j,z_j)  = \frac{1} {\epsilon_0}\frac{8}{abc} w  
	\sum_{l=1}^{N_l} \sum_{m=1}^{N_m} 
	\sum_{n=1}^{N_n} \frac{1}{\Gamma_{lmn}^2} \sin(\alpha_l x_j) 
	\sin(\beta_m y_j) \sin(\gamma_n z_j)\sin(\alpha_l x_i) \sin(\beta_m y_i) \sin(\gamma_n z_i)
\end{eqnarray}

Given particle's spatial coordinates $(x_i,y_i,\theta_i)$ in the laboratory
frame,
the particle spatial coordinates in the beam frame
$(x_i,y_i,z_i)$
can be obtained under the following approximation:
\begin{eqnarray}
	z_i & = & - l \gamma_0 \beta_0 \theta_i
\end{eqnarray}
Now, the space-charge Hamiltonian $H_2$ can be written as:
\begin{eqnarray}
H_2  = \frac{1} {2 \epsilon_0}\frac{8}{abc} w \kappa \gamma_0 \sum_{i=1}^{N_p}
       \sum_{j=1}^{N_p} \sum_{l=1}^{N_l} \sum_{m=1}^{N_m} \sum_{n=1}^{N_n}
	\frac{1}{\Gamma_{lmn}^2} \sin(\alpha_l x_j)
	\sin(\beta_m y_j) \sin(-\gamma_n l \gamma_0 \beta_0 \theta_j) \sin(\alpha_l x_i) \sin(\beta_m y_i) \sin(-\gamma_n l \gamma_0 \beta_0 \theta_i)
\end{eqnarray}
The one-step symplectic transfer map ${\mathcal M}_2$ of the particle $i$ with this Hamiltonian is given as:
\begin{eqnarray}
	p_{xi}(\tau) & = & p_{xi}(0) -
	\tau \frac{1} {\epsilon_0}\frac{8}{abc} w l \kappa \gamma_0 
	\sum_{j=1}^{N_p} \sum_{l=1}^{N_l} \sum_{m=1}^{N_m} \sum_{n=1}^{N_n}
	\frac{\alpha_l}{\Gamma_{lmn}^2} \nonumber \\
	& & \sin(\alpha_l x_j) \sin(\beta_m y_j) \sin(-\gamma_n l \gamma_0 \beta_0 \theta_j) \cos(\alpha_l x_i) \sin(\beta_m y_i)  \sin(-\gamma_n l \gamma_0 \beta_0 \theta_i) \nonumber \\
	p_{yi}(\tau) & = & p_{yi}(0) -
	\tau \frac{1} {\epsilon_0}\frac{8}{abc} w l \kappa \gamma_0 
	\sum_{j=1}^{N_p} \sum_{l=1}^{N_l} \sum_{m=1}^{N_m} \sum_{n=1}^{N_n}
	\frac{\beta_m}{\Gamma_{lmn}^2} \nonumber \\
	& & \sin(\alpha_l x_j) \sin(\beta_m y_j) \sin(-\gamma_n l \gamma_0 \beta_0 \theta_j) \sin(\alpha_l x_i) \cos(\beta_m y_i) \sin(-\gamma_n l \gamma_0 \beta_0 \theta_i)
	\nonumber  \\
	p_{ti}(\tau) & = & p_{ti}(0) +
	\tau \frac{1} {\epsilon_0}\frac{8}{abc} w l \kappa \gamma_0^2 \beta_0 
	\sum_{j=1}^{N_p} \sum_{l=1}^{N_l} \sum_{m=1}^{N_m} \sum_{n=1}^{N_n}
	\frac{\gamma_n}{\Gamma_{lmn}^2} \nonumber \\
	& & \sin(\alpha_l x_j) \sin(\beta_m y_j) \sin(-\gamma_n l \gamma_0 \beta_0 \theta_j) \sin(\alpha_l x_i) \sin(\beta_m y_i) \cos(-\gamma_n l \gamma_0 \beta_0 \theta_i)
\end{eqnarray}
where both $p_{xi}$ and $p_{yi}$ are normalized by $mC$.

\begin{figure}[!htb]
    \centering
    \includegraphics*[angle=270,width=70mm]{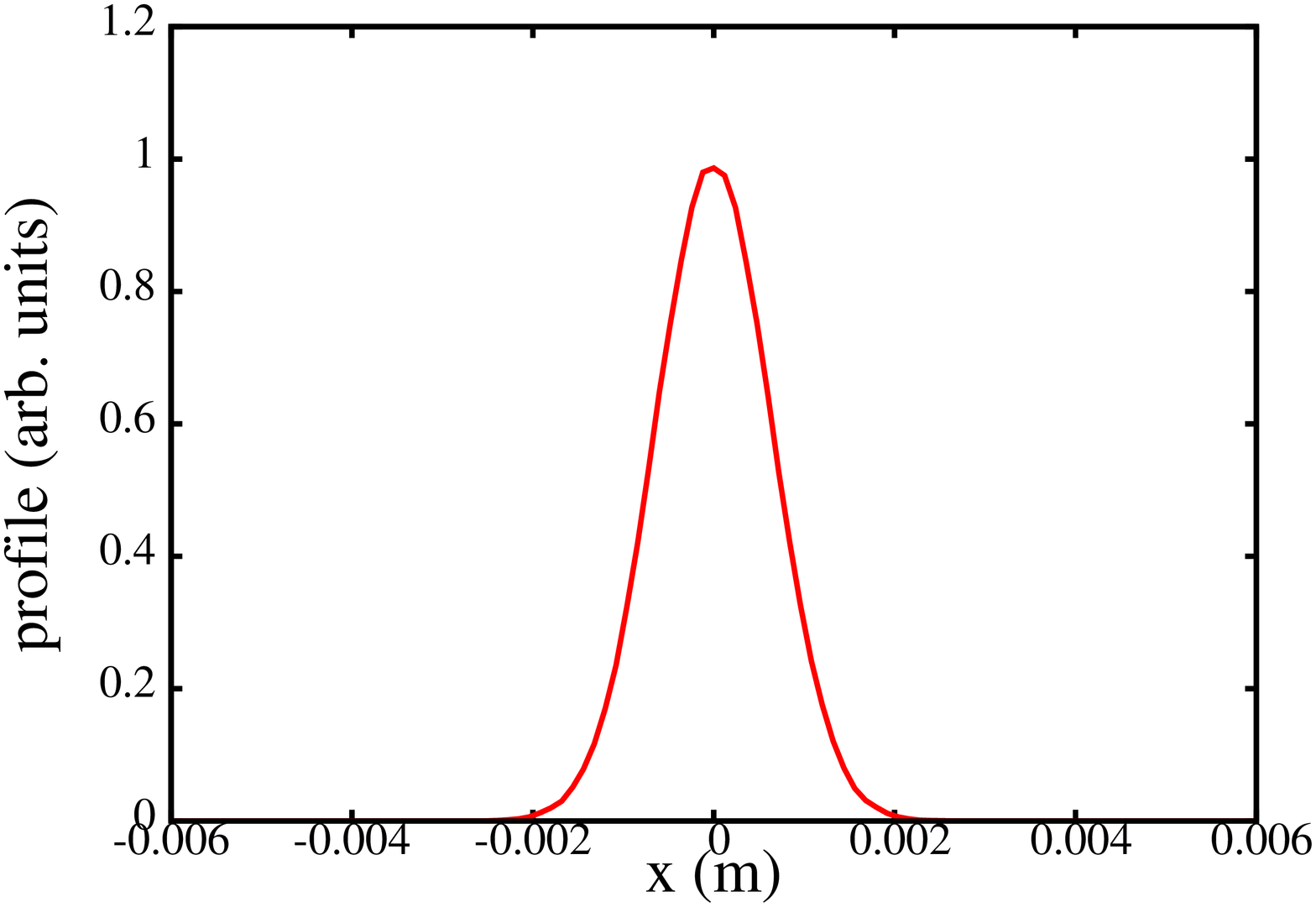}
    \includegraphics*[angle=270,width=70mm]{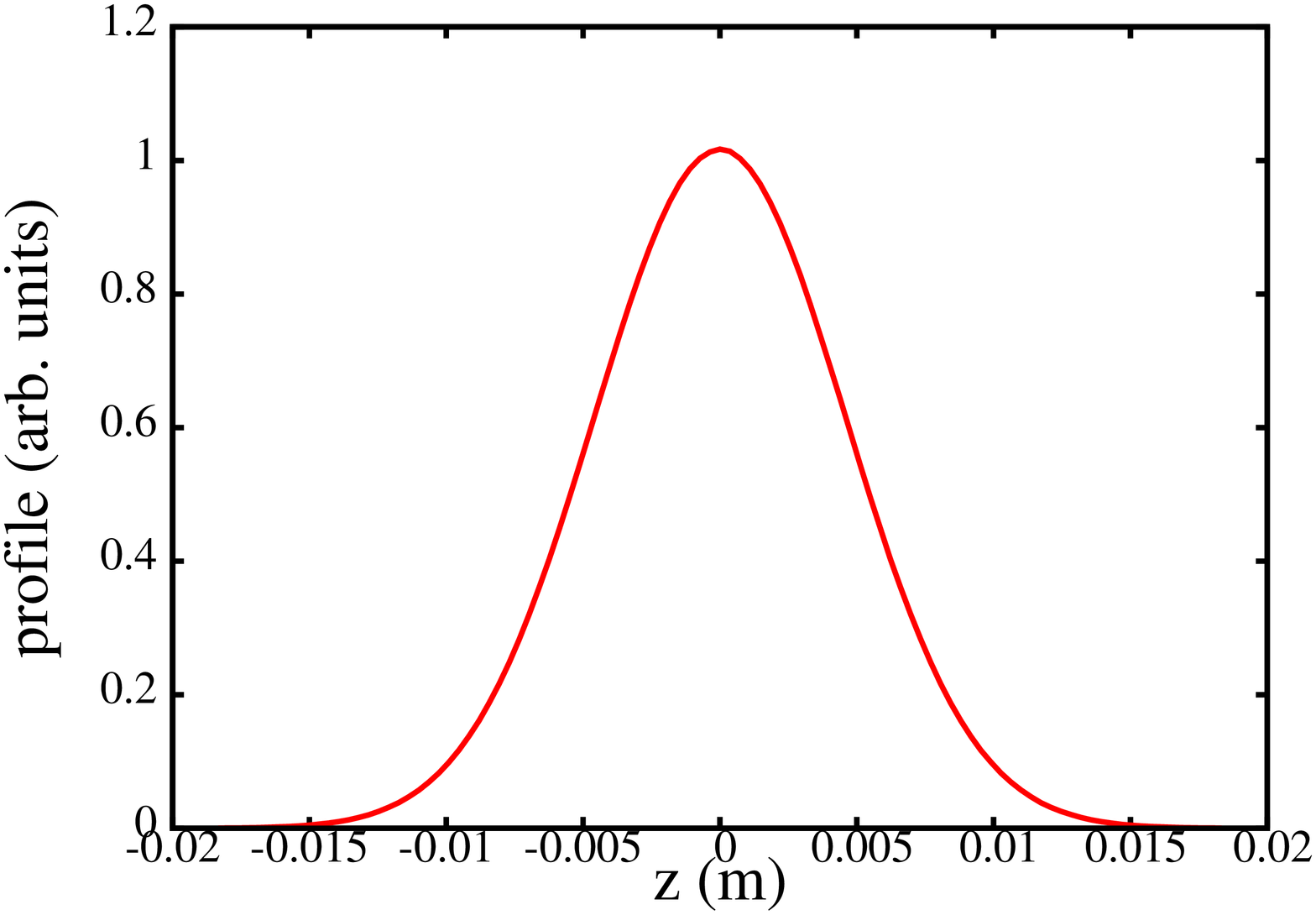}
    \caption{The transverse (left) and longitudinal (right) density profiles 
	    of a 3D bunched beam. }
    \label{prof3d}
\end{figure}

\begin{figure}[!htb]
    \centering
    \includegraphics*[angle=270,width=70mm]{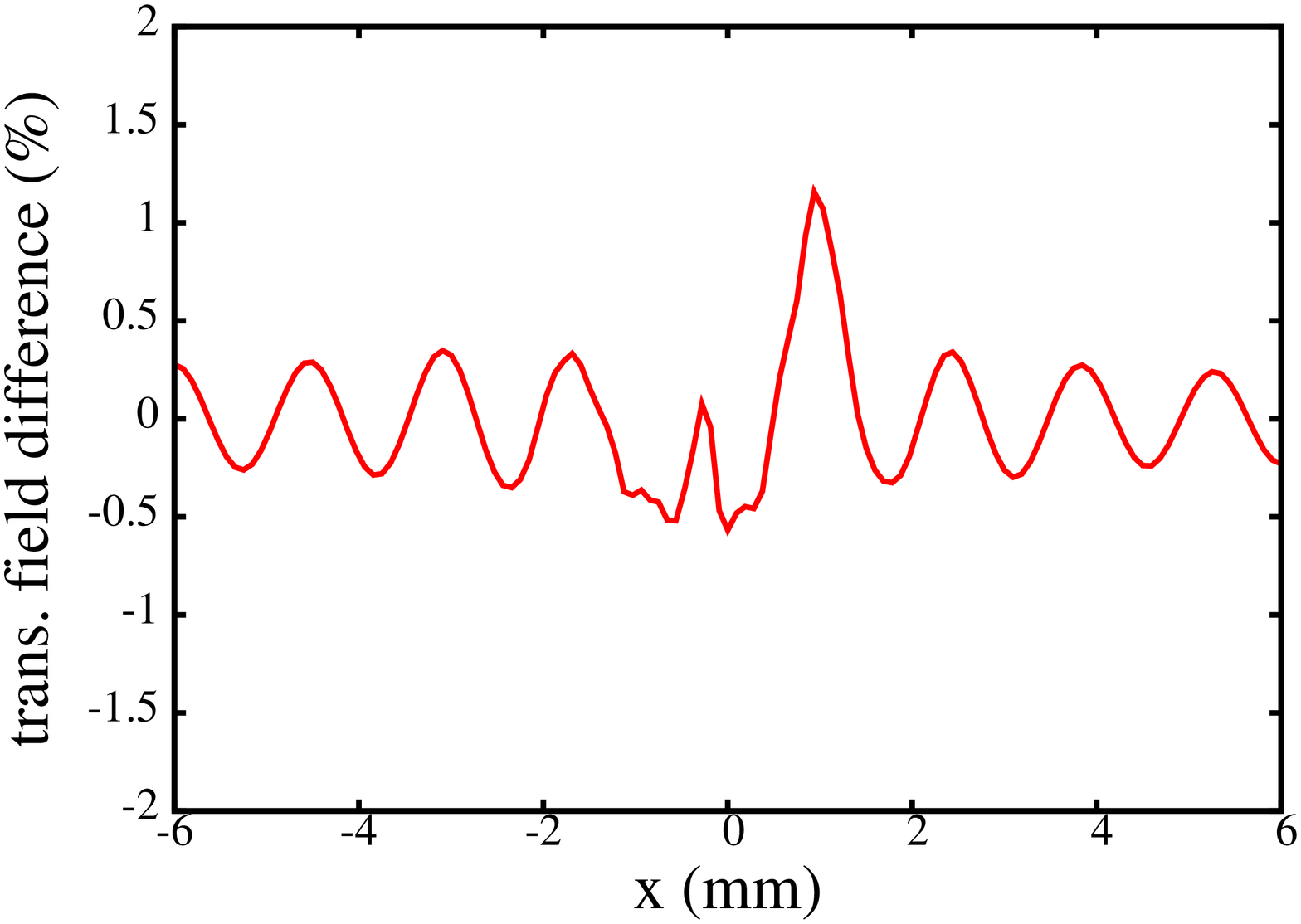}
    \includegraphics*[angle=270,width=70mm]{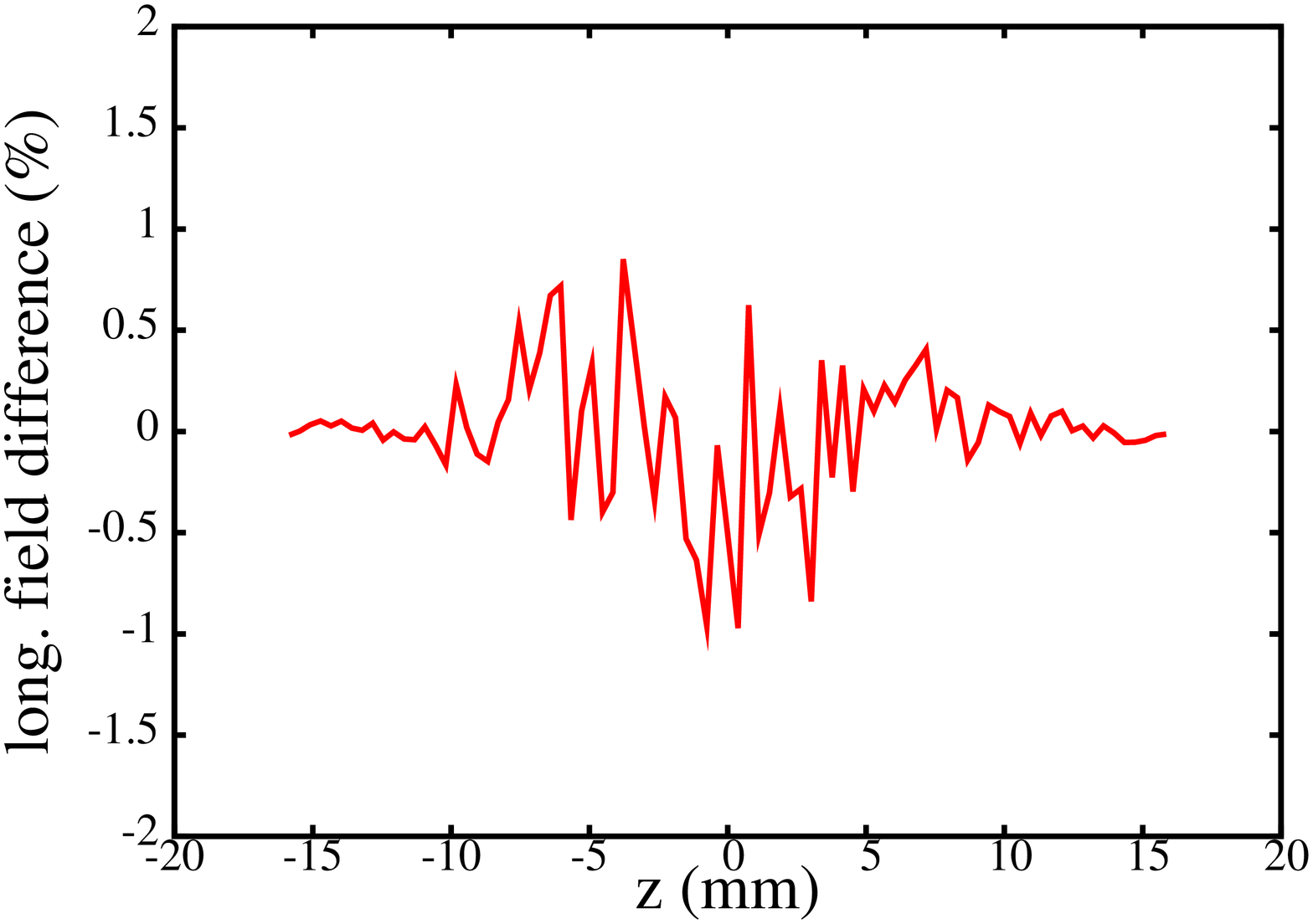}
    \caption{Relative transverse electric field difference along x-axis (left) and
	    longitudinal electric field difference along z-axis (right) from
	    the gridless spectral method and the spectral-finite
    difference solver.}
    \label{fld3d}
\end{figure}

As an illustration of above symplectic model, we simulated a $1$ GeV, 3D bunched 
proton beam transporting through a periodic focusing channel.
The initial transverse and longitudinal density profiles of the beam are
shown in Fig.~\ref{prof3d}. The beam has a 3D Gaussian distribution
with a longitudinal to transverse aspect ratio of three.
Figure~\ref{fld3d} shows the relative transverse electric field difference along the x-axis and
the longitudinal electric field difference along the z-axis using above gridless spectral
method with $15\times 15 \times 15$ modes and using a spectral-finite
difference solver~\cite{qiang1} with $129\times 129 \times 257$ grid points.
It is seen that even with only $15$ modes in each direction, the
gridless spectral solver produces space charge fields in good agreement 
with the fields from the spectral-finite difference solver with finer resolution.
The relative maximum field differences (normalized by the maximum field amplitude along the axis)
are below $2\%$ in both directions.

\begin{figure}[!htb]
    \centering
    \includegraphics*[angle=270,width=70mm]{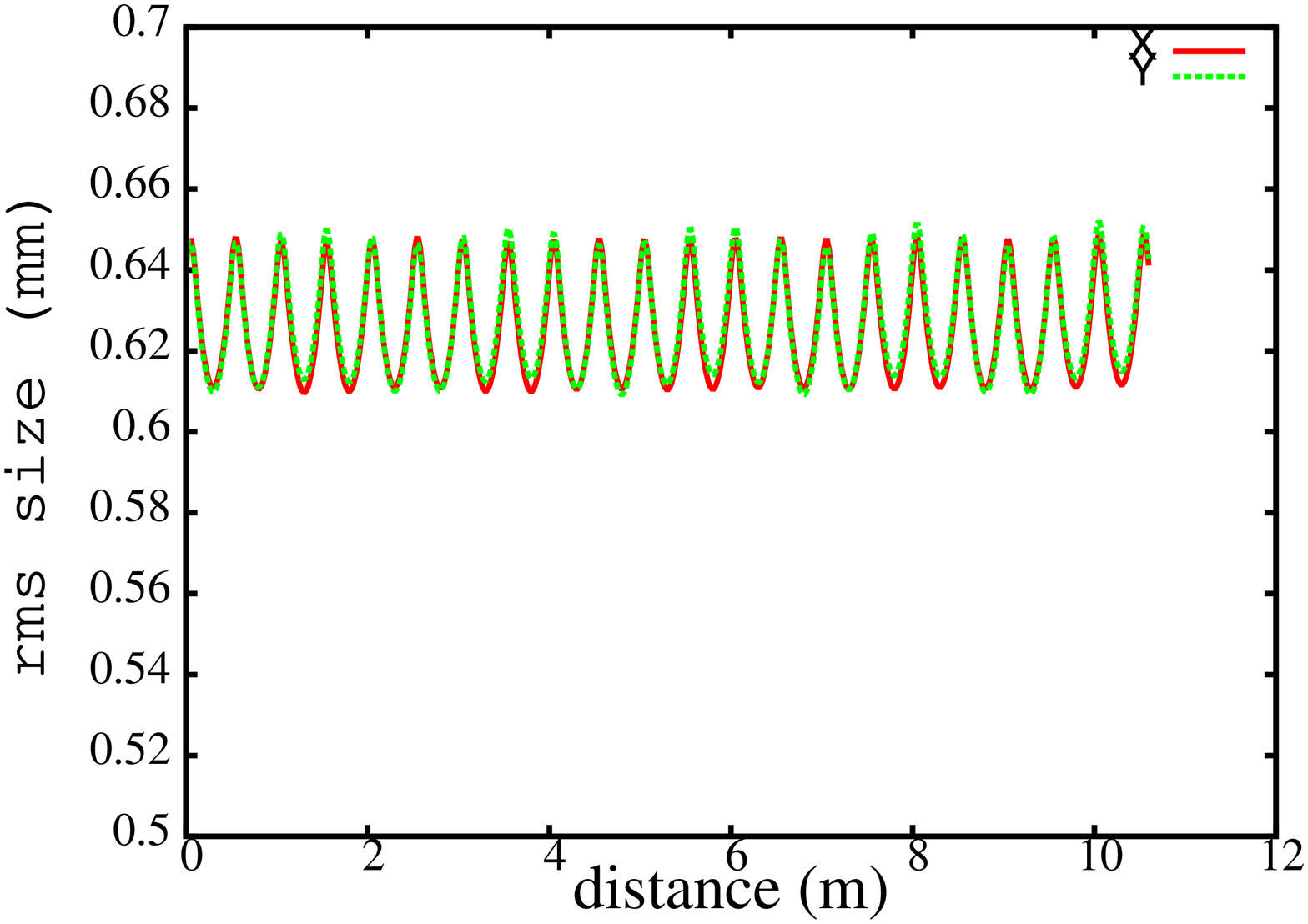}
    \includegraphics*[angle=270,width=70mm]{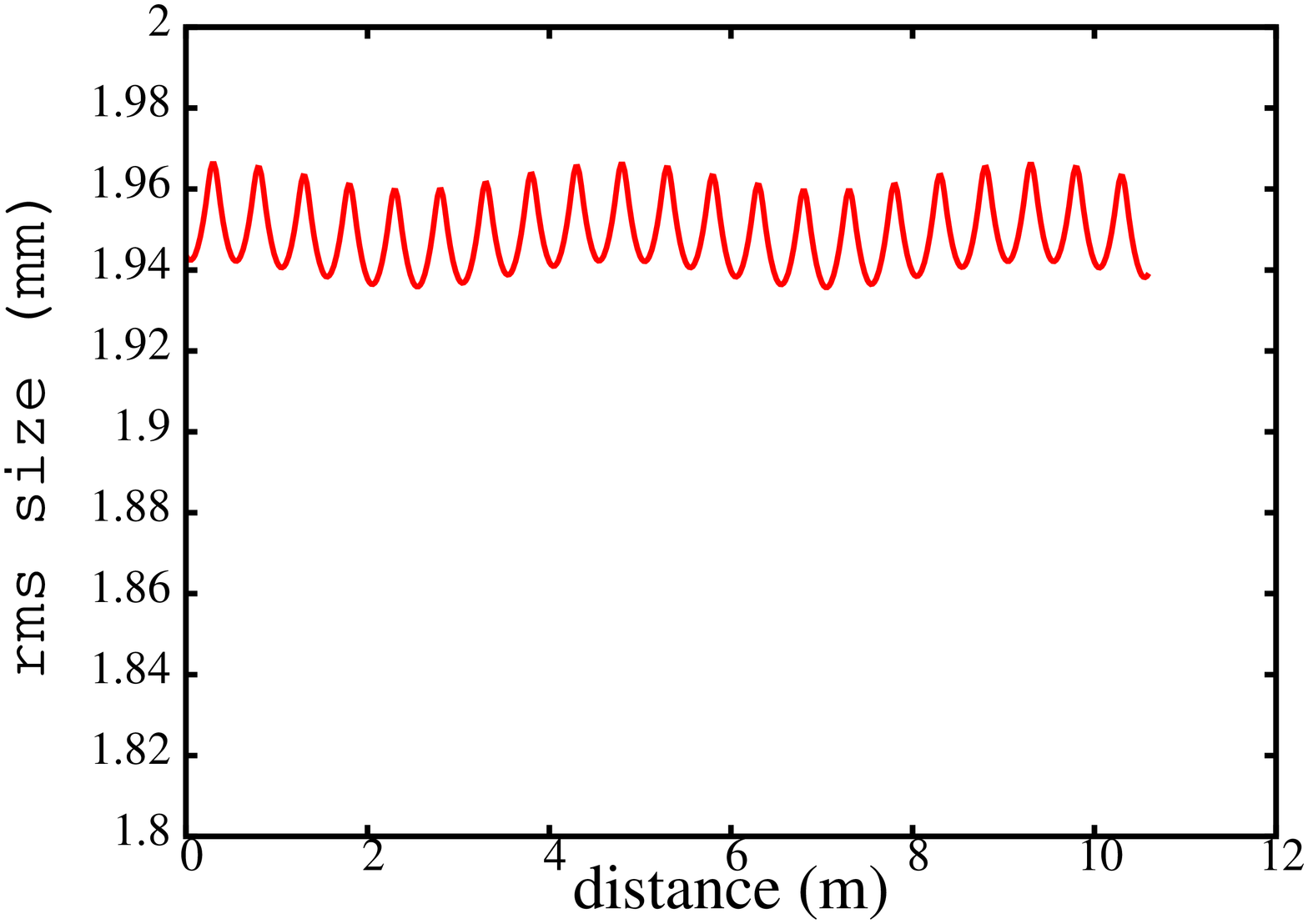}
    \caption{Transverse (left) and longitudinal (right) RMS envelope evolution
    of the bunched beam inside a periodic focusing channel.}
    \label{3drms}
\end{figure}

Figure~\ref{3drms} shows the transverse and the longitudinal rms envelope 
evolution through a periodic focusing channel.
Each period of the focusing channel consists of two transverse uniform focusing
elements, two longitudinal uniform focusing elements, and four drifts.
The total length of the period is one meter.
The zero current phase advance in a single period is about $86$ degrees in the transverse
dimension and $40$ degrees in the longitudinal direction. The phase advance
with $0.1$ A average current at $100$ MHz RF frequency
is about $81$ degrees in the transverse direction and $39$ degrees in the longitudinal direction.

\begin{figure}[!htb]
    \centering
    \includegraphics*[angle=270,width=70mm]{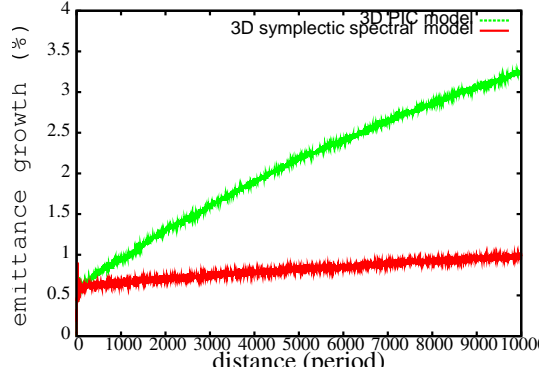}
    \caption{Six-dimensional rms emittance growth evolution in
	    the periodic channel from the symplectic spectral model (red)
    and from the PIC (green).}
    \label{6demt}
\end{figure}

Figure~\ref{6demt} shows the six-dimensional rms emittance growth 
($\frac{\epsilon_x}{\epsilon_{x0}}\frac{\epsilon_y}{\epsilon_{y0}}\frac{\epsilon_z}{\epsilon_{z0}}-1$)\%
evolution through the periodic focusing channel from the above symplectic gridless spectral model
and from the standard PIC method with the spectral-finite difference Poisson solver.
It is seen that the symplectic spectral model gives much less numerical emittance growth 
than the standard PIC method in the simulation using $160,000$ macroparticles.

\section{computational complexity}

\begin{figure}[!htb]
    \centering
    \includegraphics*[angle=270,width=70mm]{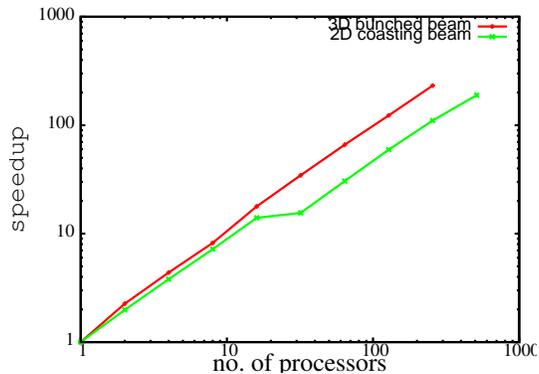}
   \caption{Parallel speedup of the 2D and the 3D symplectic tracking models on a Cray XC30 computer.}
    \label{time}
\end{figure}

The gridless symplectic multi-particle spectral model 
can be used for long-term tracking study including space-charge effects.
The computational complexity of this model scales as
$O(N_{mode} \times N_p)$, where $N_{mode}$ is the total number of modes. 
The standard PIC
model can have a computational cost of $O(N_p)+O(N_{grid}logN_{grid})$
when an efficient Poisson solver is used, where $N_{grid}$ is the total
number of grid points.
This suggests that the PIC model would be faster than the 
symplectic multi-particle spectral model on a single processor computer. 
However, the symplectic multi-particle spectral model is very 
easy to be parallelized
on multi-processor computer. One can distribute all macroparticles
uniformly across processors to achieve a perfect load balance.
By using a spectral method with exponentially decreasing errors, 
the number of modes $N_{mode}$
can be kept within a relatively small number, which significantly
improves the computing speed.
Figure~\ref{time} shows the parallel speedup of the symplectic multi-particle
spectral model as a function
of the number of processors for a fixed problem size, i.e. $\sim50,000$
macroparticles and $15\times 15$ modes in the 2D model and $\sim160,000$ macroparticles and $15\times 15 \times 15$ modes
in the 3D model.
It is seen that the speedup increases almost linearly for both models. 
This is because both models have perfect load balance among all processors.
The only communication involved in these models is a global reduction
operation to obtain the density distribution in the frequency domain.
The decrease of the speedup in the 2D case might be due to the specific 
computer 
architecture used in this timing study, which has $24$ shared memory 
computing cores inside a node. Outside the node, the communication among
processors (cores) is slowed down due to the across node communication. 
The above scaling results show that the symplectic
multi-particle spectral space-charge tracking model can have a good scalability 
on multi-processor parallel computers and is especially suitable for 
tracking simulations on large scale supercomputers or GPU computers.

\section{Conclusions}
In this paper, we proposed a new symplectic multi-particle tracking model
for self-consistent space-charge simulation.
This model uses a gridless spectral method to calculate the space-charge
potential and while avoiding the error associated with numerical grid in
the standard PIC model. It also shows much less numerical noise driven emittance
growth than the PIC method for long term simulation. 
Even though the computational cost of the symplectic
spectral model is higher than
the PIC method on a single processor computer, the proposed model
scales well on multi-processor parallel computers. It has a perfect
load balance and uniform data structure, which is suitable for GPU
parallel implementation.
The new symplectic multi-particle
spectral model enables researchers to carry out long term tracking studies including
space-charge effects.

The symplectic space-charge transfer map presented in this paper assumes
a rectangular perfect conducting pipe. A transverse open boundary condition
might be approximated using this model 
by moving the conducting wall away from the beam.
For a general boundary
condition, it is quite difficult to obtain an analytical expression of 
the electric potential from an arbitrary density distribution.
For a round perfect conducting
pipe, a Fourier mode and a Bessel mode might be used to approximate the 
particle density distribution and the electric potential of 
the Poisson equation in a cylindric coordinate
system~\cite{qiang1}. However, it takes more time to 
compute the Bessel function expansion than the simple
sine function expansion.

The symplectic space-charge model presented here also assumes a straight 
conducting pipe. 
A study of the solution of the Poisson equation in a bended conducting
pipe using the Frenet-Serret coordinate was done in reference~\cite{qiang5}
and shows that for a large normalized bending radius 
(bending radius/transverse aperture size), e.g. 100, 
there is barely any difference between the straight pipe solution and 
the bended pipe solution. 
This condition (large normalized bending radius) can be satisfied in 
most circular accelerators. Thus, the symplectic space-charge model in 
this paper can still be used for space-charge simulation in circular machines.

\section*{ACKNOWLEDGEMENTS}
Work supported by the U.S. Department of Energy under Contract No. DE-AC02-05CH11231.
We would like to thank Drs. I. Hofmann, G. Franchetti, F. Kesting, C. Mitchell, R. D. Ryne for discussions.  This research used computer resources at the National Energy Research
Scientific Computing Center.

\end{document}